# Negative differential resistance state in the free - flux - flow regime of driven vortices in a single crystal of $2H$ -NbS$_2$

### Biplab Bag[1,+], Sourav M. Karan[1], Gorky Shaw[2], A. K. Sood[3], A. K. Grover[4] & S. S. Banerjee[1,*]

[1]Department of Physics, Indian Institute of Technology, Kanpur, 208016, India

[2]National Physical Laboratory, Hampton Road, Teddington TW11 0LW, United Kingdom

[3]Department of Physics, Indian Institute of Science, Bengaluru, 560012, India

[4]Department of Applied Sciences, Punjab Engineering College, Chandigarh 160012, India

[+]Present Address: Department of Condensed Matter Physics and Materials Science, Tata Institute of Fundamental Research, Mumbai, 400005, India

*email: satyajit@iitk.ac.in

## Abstract:

Time series measurements [Gorky Shaw, *et al*. *Phys. Rev. B* **85**, 174517 (2012), Biplab Bag, et al. *Sci. Rep.* **7**, 5531 (2017)] in $2H$-NbS$_2$ crystal had unravelled a drive induced transition wherein the critical current ($I_c$) changes from a low to a high $I_c$ jammed vortex state, via a negative differential resistance (NDR) transition. Here, using multiple current ($I$) - voltage ($V$) measurement cycles, we explore the statistical nature of observing the NDR (or a quasi-NDR in reversing $I$ measurements) transition in the free-flux-flow (FF) regime in a single crystal of $2H$-NbS$_2$. Prior to the occurrence of the NDR transition, the pristine full $I$-$V$ curve exhibits a featureless smooth depinning from the low $I_c$ state. With subsequent current cycling, the NDR transition appears in the $I$-$V$ curve. Post-NDR, the full $I$-$V$ curve is seen to be noisy with depinning commencing from the higher $I_c$ state. The probability of observing the NDR transition always remains finite for a vortex state created with either fast or slow rate of magnetic field, $\dot{B}$. The probability of observing the NDR transition in the FF regime is found to systematically increase with magnetic field ($B$) in weak collective pinning regime. In the strong pinning regime, the said probability becomes field independent. Retaining of a non-zero probability for the occurrence of the NDR transition under all conditions, the observed new data shows that the $I$-$V$ branch with higher $I_c$ is the more stable compared to the lower $I_c$ branch. We show that the higher $I_c$ state, generated via the NDR transition, is unique and cannot be accessed via any conventional route, in particular, by preparing the static vortex state with a different thermomagnetic history. While the $I$-$V$ curves do not distinguish between zero field cooled (ZFC) and field cooled (FC) modes of preparing the vortex state, the probability for observing an NDR transition has different $B$-dependences for the vortex matter prepared in the ZFC and FC modes. We find that the NDR occurs in a high dissipation regime, where the flow resistivity is well above the theoretical value expected in the FF regime. We understand our results on the basis of a rapid drop in vortex viscosity at high drives in $2H$-NbS$_2$, which triggers a rapid increase in the vortex velocity and reorganization in the moving vortex matter leading to a dynamical unstable vortex flow. This dynamical instability leads to the NDR transition into a high entropy vortex state with high $I_c$.

# 1.    Introduction:

In recent times, the study of Transition Metal Dichalcogenides (TMDs), $MX_2$ ($M$ = V, Nb, Ta and $X$ = Se or S) is experiencing a renewed interest [1,2]. One reason for this interest is that these materials can be easily exfoliated down to a monolayer [3,4,5]. Therefore, in these systems, it becomes possible to explore interesting electronic properties in two dimensions which traditionally have been studied in bulk form of TMDs, like the Mott-insulating phase, superconductivity and charge density waves (CDW) state [6,7,8]. Interestingly, out of the Nb and Ta based TMDs, $2H$-NbS$_2$ is the only layered superconductor in which the superconducting gap is not modulated by an underlying CDW [9]. The $2H$-NbS$_2$ is a two-gapped superconductor with gap energies of 0.97 meV and 0.53 meV [9]. Unlike $2H$-NbSe$_2$ where a star shaped vortex core structure is found due to the influence of the CDW [10], the high resolution STM studies [9] in $2H$-NbS$_2$ show that the vortex cores in it are circular shaped with quasi-particle bound states within a core. Recently, it has also been shown for doped $2H$-NbSe$_2$ that magnetic impurities affect the nature of bound states inside the vortex core [11]. All of these suggest that it is worthwhile asking if some of the above peculiar features of $2H$-NbS$_2$ affect the nature of the driven vortex state in this system. Unfortunately, only a few recent studies exist on the nature of driven vortex state in $2H$-NbS$_2$.

A current ($I$) - voltage ($V$) measurement is equivalent to measuring the force ($F$) versus the average vortex velocity ($u$) behaviour for the driven vortex state in type II superconductors, where $F$ ($\propto I$) is the Lorentz force driving the vortices and $u = V/_{Bd}$ ($d$ is the separation between the $V$-contacts). The dynamics of driven vortices through a random pinning environment serve as a prototype for diverse hard and soft condensed matter systems [12,13,14,15,16,17,18,19,20]. Typically, when the applied current density exceeds the critical depinning current density [21,22,23,24], vortices in the superconductor begin to move, thereby inducing a voltage. In the $I$-$V$ curve of $2H$-NbS$_2$ [25], one identifies a thermally-activated-flux-flow (TAFF) regime below a current, $I_{cr}$. Beyond a break in curvature (at $I_{cr}$) in $I$-$V$, the driven vortex state enters the free-flux-flow (FF) regime with a linear $I$-$V$ response. Studies in recent times showed that at field $B$, the $I$-$V$ of a vortex state in $2H$-NbS$_2$ exhibits a drive induced transformation from a conventional low $I_c$ ($I_c^l$) state to a high $I_c$ ($I_c^h$) state [26,27]. The transformation is characterised by a sharp negative differential resistance (NDR) transition, which corresponds to negative $dV/_{dI}$, i.e., a decrease in $V$ ($\equiv u$) when $I$ is increased. Note that in our paper the NDR refers to a sharp drop in $V$ while increasing $I$, and a quasi-NDR refers to a sharp drop in $V$ observed while reducing $I$ in an $I$-$V$ measurement. The $I_c^h$ state was considered to be a drive-induced state and was identified as a jammed vortex state. Measurement of time ($t$) series of large $V$ fluctuations ($V(t)$) revealed unusual features near $I_c^h$, viz., (a) a critical behaviour like divergence in time scale for which the fluctuations are sustained [26], with a divergence exponent reminiscent of random organization in a driven colloidal matter [28], and (b) the fluctuations obey a popular Gallavotti-Cohen non-equilibrium fluctuation relation [27,29]. These studies implied that the jammed state was quite distinct from the conventionally prepared pinned static vortex matter, either in field-cooled or zero-field cooled.

In $2H$-NbS$_2$ crystals, $V(B)$ measurements at constant $I$ as well as $I$-$V$ measurements at constant $B$ [26,27] showed that the static vortex matter prepared using ZFC mode, in which $B$ was reached using a field sweep rate ($\dot{B}$) < 0.02 T/min, depinned from a high $I_c$ state. However, for $\dot{B} >> 0.02$ T/min, the ZFC vortex state (for the same $B$) depinned from a low $I_c$ state. For $\dot{B} \approx 0.02$ T/min, depinning commenced at $I_c^l$. However, at higher drives, the moving vortex matter abruptly came to a halt and transformed into the jammed state with higher $I_c$ ($I_c^h$). It may be mentioned in these earlier studies [26,27], there was no detailed exploration of the nature of $I$-$V$ curves for the field cooled (FC) vortex matter. In these earlier studies, one only verified that by waiting long enough in the $V(t)$ measurements, the driven FC state at 0.8 T transforms back to a high $I_c$ state, which exhibits similar features in the $V(t)$ measurements as those found for the ZFC state. A detailed investigation of the conditions under which the unusual NDR transition occurs and its dependence on $B$, $\dot{B}$ and the thermomagnetic history of the starting static vortex state, has been lacking.

It is pertinent to recall here a few general features known from studies on the vortex state. In realistic samples with moderate to strong pinning, the $I_c$ of the static ZFC vortex state is usually lower [30, 31, 32] than that for the FC state [30-32], as during field cooling, the pinning centers freeze in a more spatially disordered pinned vortex configuration leading to a higher $I_c$. Only in very clean samples, the $I_c$ of the ZFC and FC states are undiscernibly close. Usually, a high $\dot{B}$ results in annealing of disordered vortex state into an ordered state with low $I_c$ [30,31,32,33]. The glassy vortex state exhibits memory and metastability effects [31,32]. Quite a few of these features have been explained via an edge contamination process [34,35], an ever-present process in the driven (with transport current) vortex state. In $I$-$V$ measurement, there is a continuous injection of disordered phase from irregular sample boundaries into the vortex state in the sample at a field $B$ and temperature $T$. Within this process, a low $I_c$ is favoured, when the rate at which the disordered phase is injected from the irregular sample edges is slower than the rate at which the injected disordered phase anneals, as it mixes with a ordered vortex phase already present inside the sample. If the order of the above rates reverses, then a disordered or an admixed vortex phase is favoured. The annealing rate also depends on the ($B$, $T$) of the static vortex state, as the annealing rate is faster for an elastic vortex solid, while it is slower for a softened vortex state. In the backdrop of these features, it may be worth to further investigate the $I$-$V$ features in $2H$-NbS$_2$.

Using high sensitive dc magneto-transport measurements in a $2H$-NbS$_2$ single crystal, we explore the statistics of observing the NDR transition in the FF regime and study its dependence on $B$, $\dot{B}$ and the thermomagnetic history of the static vortex state. We find that the $I$-$V$ curves are identical for the FC and ZFC vortex states. We examine features developing in the $I$-$V$ with repeated current cycling between zero and high current value (without causing heating at the contacts). Our study shows that the NDR (or a quasi-NDR) transition may not appear in the first $I$-$V$ run, and a few $I$ cycles are required to generate the NDR transition in the FF regime. These studies determine the $B$-dependence on the number of the current cycles ($N$) needed to generate an NDR (or a quasi - NDR) transition. The $N$ is inversely related to the probability of observing an NDR (or quasi-NDR) transition. We find statistically that this probability decreases with an increase in $B$. We also find that irrespective of the value of $\dot{B}$ value, with a sufficient number of $I$ cyclings, there is always a non-zero probability for observing an NDR (or quasi-NDR) transformation. The above results cannot be fully reconciled within the edge contamination process alone. While the $I$-$V$ curves exhibit identical features for the vortex matter prepared via ZFC and FC modes, the $N(B)$ shows distinct behaviour for the ZFC and FC curves. The difference in ZFC−FC behaviour of $N(B)$ is related with a crossover from strong to weak pinning regime of the underlying static vortex solid. We discuss our results in terms of dynamical instability in the vortex flow present in $2H$-NbS$_2$. The dynamical instability triggers the NDR transition into an unusually high entropy vortex configuration with a high $I_c$.

## 2. Experimental details:

We perform four-probe transport measurements in a single crystal of $2H$-NbS$_2$ superconductor (approximate dimension of $1.9 \times 1.0 \times 0.045$ mm$^3$) by passing current in the basal ($ab$) plane of the single crystal with $B$ applied along the crystallographic $c$ axis ($B\|c$). In the section 1 of Supplemental Material [36], we also show results on another crystal (A2) from different batch with dimensions $0.9 \times 0.9 \times 0.045$ mm$^3$. The single crystals were grown using the chemical vapour transport technique with growth details reported elsewhere [37]. The powder XRD analysis of $2H$-NbS$_2$ samples from different batches as well as the energy dispersive X-ray analysis (EDX) study are presented in section 2 of Supplemental Material [36]. The study shows the presence of only the $2H$-NbS$_2$ superconducting phase. The zero-field superconducting transition temperature ($T_c$) and the residual resistivity ratio [RRR $\equiv R(300$ K$)/R(10$ K$)$] of the present sample are $5.8 \pm 0.1$ K and 25, respectively (In crystal A2, $T_c$ and RRR are 5.8 K and 35). Four probe electrical contacts are made on a freshly cleaved sample surface using low-temperature silver epoxy with a mean spacing between the voltage contacts, $d = 0.55 \pm 0.01$ mm (contact resistance $\sim 10 \pm 4$ m$\Omega$). The $I_c$ is determined from $I$-$V$ measurements using a voltage criterion, viz., $V \geq 2$ $\mu$V at $I = I_c$. Before the $I$-$V$ measurements, the pristine static vortex matter is prepared via either ZFC or FC mode at a field $B$. The $I$-$V$ of different ZFC vortex states, prepared by reaching the target field $B$ using different field sweep rates $\dot{B}$, are studied. In our paper we have studied different field sweep rates. A field sweep rate of 0.1 T/min, corresponds to induced emf of the order of 3.2 nV. A current of 10 mA (for the crystal whose results are shown in the paper) corresponds to a current density of $2.2 \times 10^5$ A.m$^{-2}$ (note that due to the nearly comparable cross-section of both the crystals whose data are shown in the paper and in Supplemental Material [36], the current densities for both crystals are nearly similar). However, we would like to mention here that we do not record any $I$-$V$ measurements while the field is being swept towards the target field. After reaching the target field value, the field is stabilized with a wait time of up to 5 minutes. After this wait time, we perform the $I$-$V$ measurement. Hence, the emf of 3.2 nV induced by field sweeping is a transient which does not affect our $I$-$V$ measurements. Furthermore, these transient voltages are far smaller than the voltages we measure.

The *I-V* is recorded while driving the vortices by ramping up (or down) the dc applied current. For all *I-V* measurements, the *I* sweep rate was ~ 10 mA/min. The *I-V* measurement for the forward (increasing *I*) and reverse (decreasing *I*) runs labelled as F# and R#, where '#' is a number referring to the leg of the current cycle (see subsequent text). The present crystal is from the same batch as the A1 crystal on which time-series measurements related to the jamming transition had been reported in Ref. [27].

## 3. Results:

### 3.1 Generating the NDR transition in *I-V* and the field dependence of high critical current state

Our earlier measurements [26] on another crystal of 2*H*-NbS$_2$ had explored the jamming transition, which was an abrupt jump to a low *V* state, as seen in an isothermal *V-B* measurement. In Ref. [26], using *I-V* measurements it had shown that the transition was associated with a field sweep rate-dependent transformation into a high *I$_c$* state. The work, however focussed on characterizing the nature of the non-equilibrium jamming transition via detailed time-series voltage measurements [26,27]. The instabilities in *I-V* characteristics due to jamming are explored in greater detail in this paper. Figure 1(a) shows the *I-V* response in log-linear scale for a ZFC vortex state prepared at *T* = 2.5 K and 0.3 T with $\dot{B}$ = 0.03 T/min. It may be recalled that earlier studies in the sample A1 suggested that in *I-V* for $\dot{B}$ > 0.02 T/min, no NDR transition into the high *I$_c$* state was seen (see Fig. 2 of Ref. [26]). In our subsequent measurements in the present sample here, we label the first forward *I-V* measurement as the F1 run, where *I* is being increased from zero up to 100 mA. We would like to mention that in the F1 measurements, we observe that whether we prepare the static vortex state in ZFC or FC mode, the measured depinning threshold of the pristine static state ($I_c^I$) for both these modes are identical, and the corresponding *I-V* curves overlap. We identify this *I-V* curve as the low *I$_c$* branch of *I-V*. The *y*-axis on the right-hand side of Fig. 1(a) presents the corresponding *u(I)* values determined from the *I-V*. As *I* is increased from zero in the F1 run, the vortex state depins at $I_c^I$ ~ 23 mA. After depinning, the moving vortex matter exhibits a linear TAFF regime. The linear behaviour in the log-linear scale suggests an Arrhenius-like thermally activated motion [25] until the knee in the *I-V* is reached, *viz.*, at *I$_{cr}$* ~ 34 mA.

In Fig. 1(a), the *I-V* characteristic for F1 has a break in slope at *I$_{cr}$*, which corresponds to a crossover from TAFF into the free-flux-flow (FF) regime [25]. We observe a monotonic increase in *u* with drive in this FF regime as *I* is ramped up to 100 mA. Typically, the FF regime is considered a uniform vortex flow regime. The low *I$_c$* *I-V* branch, in this FF regime, is completely reversible while increasing (F1) and decreasing (R1) the current between *I$_{cr}$* and 100 mA. During R1 (*I*: 100 mA → 0 mA), the *I-V* curve completely overlaps with the forward one (F1) run (see Fig. 1(a)). Note, the *V(I)* falls below the noise level (< 2 μV) at $I_c^I$ ~ 23 mA for R1 run as well. The complete overlap between the F1 and R1 curves in Fig. 1(a) rules out any significant Joule heating effect at contacts.

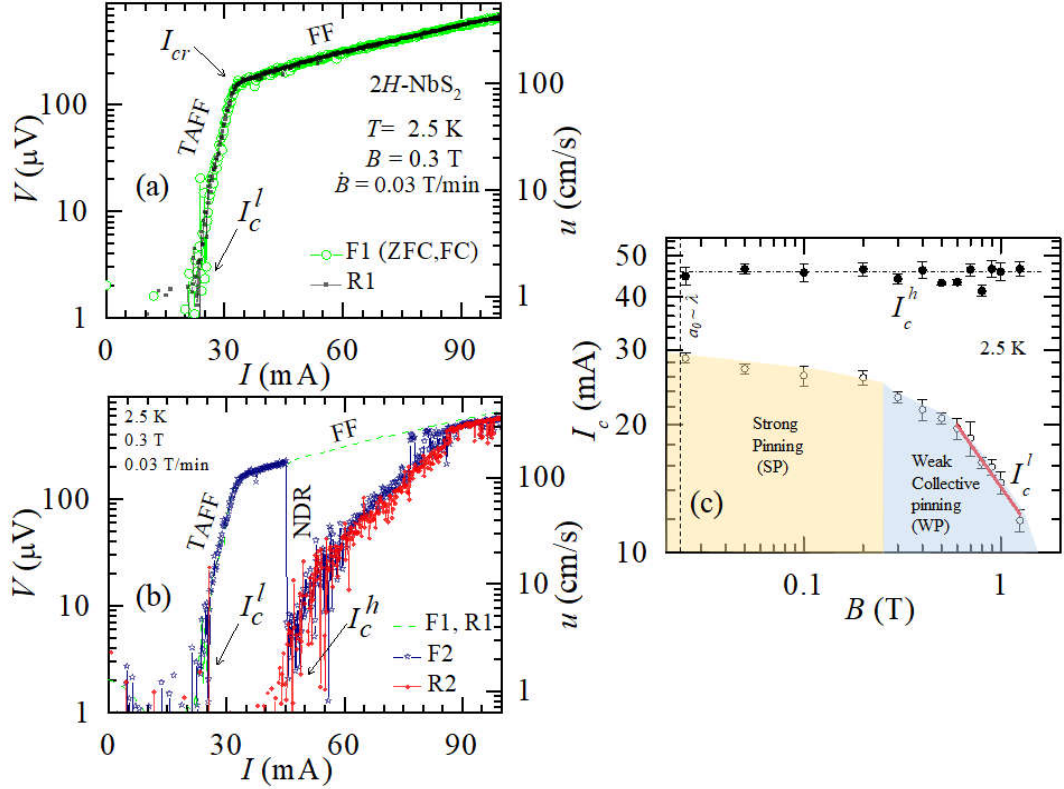

Figure 1. The left and right *y*-axes show the voltage, *V* and the corresponding vortex velocity *u* data versus *I* (plotted in a log-linear scale) for the driven vortex state at 2.5 K and 0.3 T for (a) F1 and R1 (see text for details), & (b) F1, R1, F2 and R2 runs. The pristine vortex state at $B = 0.3$ T is reached with a $\dot{B} = 0.03$ T/min. In (a), the F1 curve represents the *I-V* response for ZFC and FC pristine static vortex matter. In (b), the green dashed line represents the F1 runs of (a). The locations of $I_c^l$, $I_{cr}$ and $I_c^h$ are marked with arrows in (a) and (b). (b) shows the NDR transition into the high $I_c$ state in the F2 R2 run. (c) show the *B*-dependence of $I_c^l$ and $I_c^h$ estimated from isothermal *I-V*'s at 2.5 K. The thick red line presents $I_c^l \propto B^{-\alpha}$ behaviour with $\alpha = 0.6 \pm 0.1$. The yellow and blue regions identify strong pinning (SP) regime and weak collective pinning (WP) regime, respectively. The vertical dashed line represents the $a_0 \sim \lambda_{ab}$ line.

In Fig. 1(b), the 2$^{nd}$ *I-V* cycle F2 (viz., *I* is again increased from zero towards 100 mA) exhibits features that are similar to F1 and the vortex state depins at $I_c^l \sim 23$ mA. However, for the F2 run above $I_{cr}$, as the driven vortex state enters the FF regime (Fig. 1(b)), *I-V* exhibits an abrupt NDR transition at $I \sim 45$ mA and the average vortex velocity abruptly drops down from 100 cm/s (in the FF regime) to 4 cm/s. With the subsequent increase of *I* beyond 45 mA, the *V(I)* increases gradually, however, there is a lot more voltage noise compared to the smooth *I-V* while depinning at $I_c^l$. The large *V(I)* or *u(I)* fluctuations imply instability in vortex flow beyond the NDR transition at 45 mA. It appears that between ~ 45 mA and ~ 80 mA, the *I-V* curve along the FF branch in the F1 run is unstable. Beyond 80 mA, the fluctuating *I-V* in F2 merges with that of the FF regime (Fig. 1(b)). Upon reducing current down from 100 mA (R2 run), *I-V* response displays similar *V* fluctuations as in the F2 run. Also, the *V(I)* for the R2 run falls below the noise level (< 2 µV) at $I \le 44$ mA ($\equiv I_c^h$), indicating driven vortices enter the high $I_c$ state. It is important to note that after the onset of this high $I_c$ vortex state in R2 run, the driven vortex matter in subsequent $I$ cycles ($0 \rightarrow 100 \rightarrow 0$ mA→ ….) depins only from the high $I_c$ state, and depinning from the low $I_c$ state (at $I_c^I$, observed for the pristine F1 run) is never observed again. We call this the high $I_c$ branch of the $I$-$V$ curve. Thus, with sufficient number of current cycling, the NDR is always encountered in the FF regime with $\dot{B} = 0.03$ T/min, and the $I$-$V$ curve displays irreversible behaviour with two branches of the $I$-$V$ curve (one being the low $I_c$ and the other is the high $I_c$ branch of the $I$-$V$ curve).

In Fig. 1(c), we find that with increasing $B$, $I_c^I(B)$ varies as $1/B^\alpha$, where $\alpha = 0.6 \pm 0.1$ for the red line in Fig. 1(c). This form of the $I_c^I(B)$ represents the weak collective pinning regime of the elastic vortex matter [21,22,24]. Within the region shaded light blue in Fig. 1(c), $I_c^I(B)$ $\propto 1/B^\alpha$. In Fig. 1(c), we show the dashed vertical as the field values where the intervortex spacing, $a_0 \sim \sqrt{\frac{\phi_0}{B}} = \lambda_{ab}$ (where $\phi_0$ is the magnetic flux quantum and $\lambda_{ab}$ is $ab$ plane penetration depth), and using $\lambda_{ab} \sim 350$ nm in $2H$-NbS$_2$ [38], we get a $B \sim 0.016$ T. At low fields, viz., inside the light yellow regime in Fig. 1(c) (which is near $a_0 \sim \lambda_{ab}$ line), due to relatively weak inter-vortex interaction, the vortex state is in a strong pinning regime. In the strong pinning regime, the pinning force almost saturates, and hence $I_c^I$ has a weak field dependence in this regime. In Fig. 1(c), we note that the high $I_c$ state has almost no $B$ dependence and the $I_c^h$ value is higher than the maximum strong pinning regime value. We would like to mention that presence of any strong pinning centers like structural imperfections in the crystal or compositional variations will affect $I_c^l$. Had these strong pinning centres played a role in reaching the high $I_c$ state, then we should have observed this high $I_c$ upon field cooling the sample. In the section 3 of Supplemental Material [36], we show that by field cooling we do not obtain the high $I_c$ state, rather we get the low $I_c$ state.

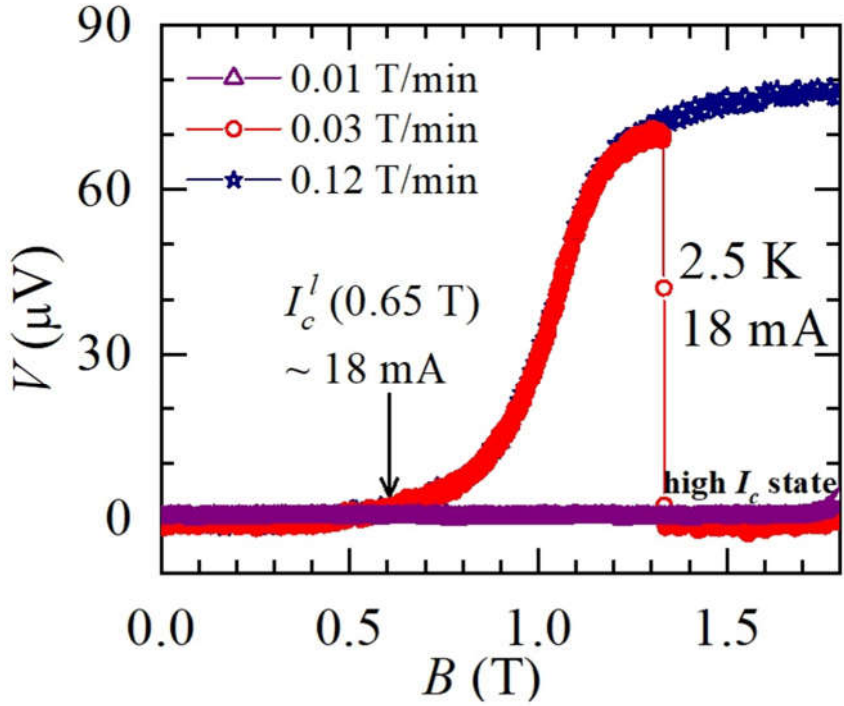

Figure 2. $V(B)$ data at 2.5 K and a constant $I = 18$ mA for $\dot{B} = 0.01$, 0.03 and 0.12 T/min. For data taken with high field sweep rate of 0.1 T/min, the $V(B)$ shows depinning ($V(B) > 2$ μV) at a $B$ value when the $I_c(B)$ is less than 18 mA. For field sweep rate of 0.03 T/min, while the vortex state depins from a low $B$, it subsequently transforms into a high $I_c$ state at higher $B$. With a slower field sweep rate of 0.01 T/min, the vortex state never depins as it remains in the high $I_c$ state (confirmed with $I$-$V$ measurements).

## 3.2 **Effect of magnetic field ramp rate on the generation of high $I_c$ state**

Figure 2 displays $V$ versus $B$ behaviour measured with a constant $I$ = 18 mA at 2.5 K for different $\dot{B}$ values. The observed jump in $V(B)$ and the $\dot{B}$ dependence in $V(B)$ measurement in this crystal of 2$H$-NbS$_2$ is similar to the $V(B)$ behaviour associated with the jamming transition shown in Fig. 1 of Ref. [26], in a different crystal of 2$H$-NbS$_2$. Figure 2 suggests that the jamming transition feature seen in $V(B)$ measurement corresponds with the NDR transition found in $I$-$V$ measurements. From Fig. 2 we see that for a ZFC vortex state prepared using a high $\dot{B}$ = 0.12 T/min, the vortex state depins from a low $I_c$ value, viz., the $B$ at which $I_c^l(B) <$ 18 mA, depinning occurs. However, for vortex states prepared up to 1.8 T with low $\dot{B}$ = 0.01 T/min, the vortex matter does not get depinned in response to drive with $I$ = 18 mA, thereby implying that the underlying vortex matter does not have a low $I_c$. This feature is consistent with the results of Ref. [26], which showed that for $\dot{B}$ < 0.02 T/min, the vortex matter attains the high $I_c$ state (see inset of Fig. 1 in [26]). From the $V(B)$ response for $\dot{B}$ = 0.01 T/min in section 4 [36] for the *I-V* data). In other words, with $\dot{B} = 0.03$ T/min followed by depinning at $I_c^l(B)$, at higher $B$ there is an NDR transition in $V(B)$, where the vortex matter re-organizes into the high $I_c$ state.

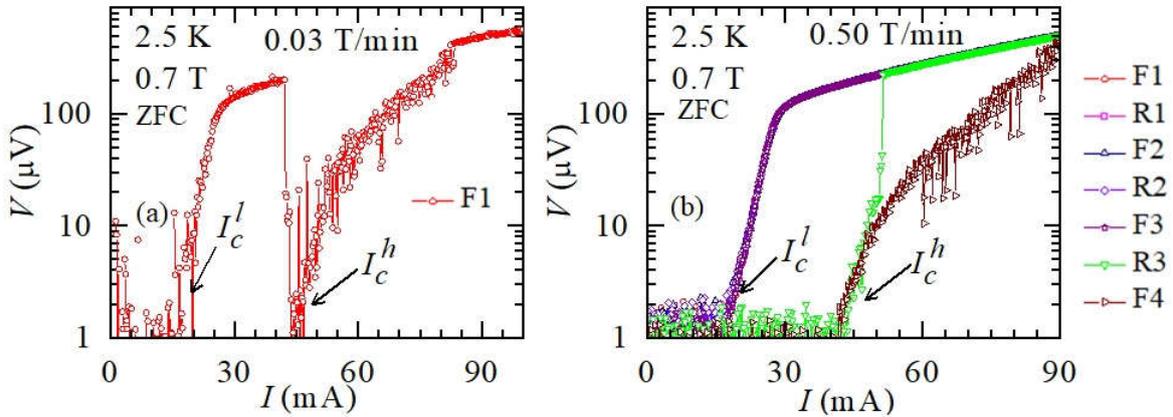

Figure 3. $I$-$V$ data (for F1, R1, F2, R2, F3, R3 and F4 runs) at 2.5 K and 0.7 T (ZFC) for (a) $\dot{B}$ = 0.03 T/min and (b) 0.50 T/min, respectively. The locations of $I_c^l$ and $I_c^h$ are marked with arrows. (see text for details)

Fig. 2, it is clear that the vortex state remains pinned with the $V$ values staying in the noise floor, indicating the vortex state has been prepared in a high $I_c$ state (i.e., $I_c$ > 18 mA for $B$ < 1.8 T at 2.5 K). However, at an intermediate sweep rate of 0.03 T/min, the vortex matter gets depinned from a low $I_c$ value and subsequently it exhibits an NDR transition at $B \sim$ 1.32 T, where the voltage falls sharply below the noise level. Here, the vortex state makes a transition into the high $I_c$ state. Following the NDR drop at $B$ > 1.35 T with $\dot{B}$ = 0.03 T/min, a measurement of $I$-$V$ shows vortex state depinning at $I_c^h \sim$ 45 mA (see Supplemental Material,

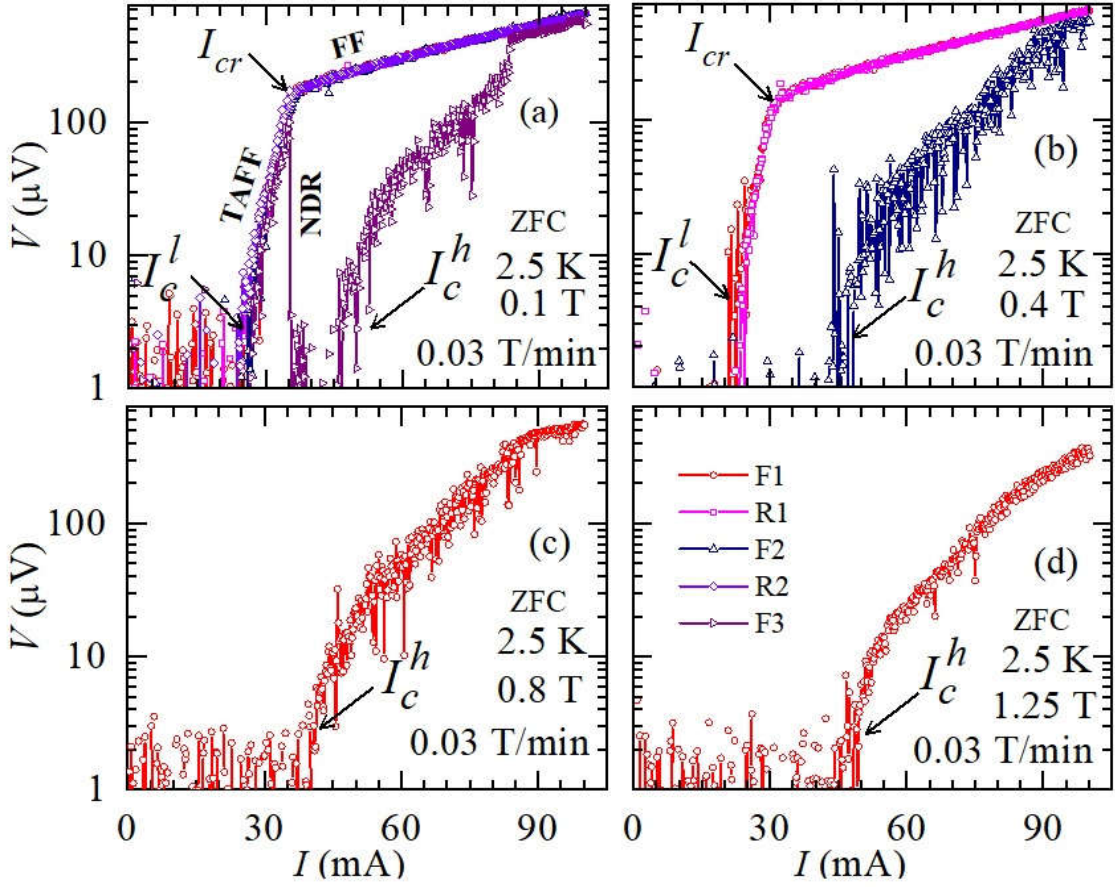

Figure 4. *I-V* data of the driven vortex matter at 2.5 K and (a) 0.01 T, (b) 0.04 T, (c) 0.8 T and (d) 1.25 T. The pristine vortex state is prepared in ZFC mode at $\dot{B} = 0.03$ T/min. $I_c^l$ and $I_c^h$ locations are marked with arrows. The current cycle legends shown in (d) are retained the same for (a)-(d).

Figures 3(a), 3(b) show the *I-V* data for a ZFC static vortex state prepared at 2.5 K and 0.7 T with $\dot{B} = 0.03$ T/min and 0.5 T/min, respectively. Figure 3(a) shows that for the F1 run after depinning from a low $I_c^l \sim 17$ mA, the *I-V* exhibits an NDR transition at 41 mA (from the FF regime), and the driven vortex matter enters the high $I_c$ state ($I_c^h \sim 45$ mA). By comparing with Fig. 1 ($B = 0.3$ T), it appears that for the same $\dot{B}$, a fewer number of current cycles is required to observe the NDR transition at higher $B$ values, viz., the probability for observing the NDR transition in the *I-V* increases with $B$. When the same $B = 0.7$ T is reached with a very high $\dot{B}$ = 0.5 T/min, which is almost 25 times the value of 0.02 T/min of Ref. [26] below which there is a transformation from the low to high $I_c$ state, one had not expected to observe any NDR transition into a high $I_c$ state. However, the vortex state prepared at 2.5 K and 0.7 T with $\dot{B} = 0.5$ T/min, in Fig. 3(b) shows a depinning from the low $I_c^l$ state. The *I-V* curves remain reversible over multiple *I* cycles initially. This reversible feature is survived only until the 3[rd]

forward run (F3). In the third reverse run (R3), the high $I_c$ state resurfaces via the quasi-NDR feature. After this quasi-NDR transition, upon reducing the $I$ to zero and in the subsequent F4 run, the depinning occurs at $I_c^h \sim 45$ mA. It is clear that with a sufficient number of $I$ cycling's, the higher $I_c$ branch of the $I$-$V$ curve will be reached via an NDR (or quasi-NDR) transition.

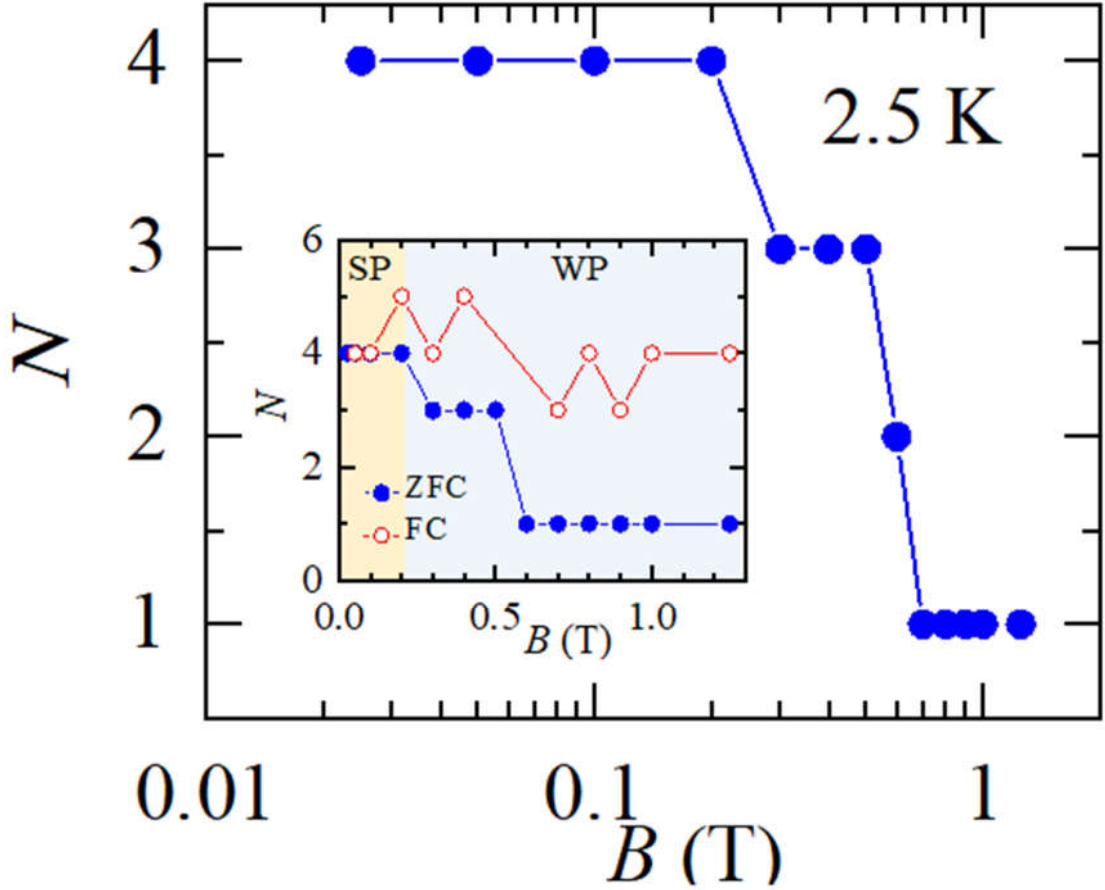

Figure 5. Main panel shows the number of $I$ cycles needed for the $I$-$V$ to show the NDR transition, i.e., $N(B)$ at 2.5 K (for the static vortex state prepared in the ZFC mode). Inset shows comparison of $N(B)$ values for static vortex state prepared in the FC and ZFC modes at 2.5 K. In the inset, the yellow and blue regions represent the SP and WP regimes (compared with Fig. 1(c)), respectively.

### 3.3 Variations of depinning characteristics across $I_c^l$ and $I_c^h$ with $B$ and the $N(B)$ plot

Figures 4(a)-4(d) show the $I$-$V$ of the driven vortex state at $T = 2.5$ K for $B = 0.1$, 0.4, 0.8 & 1.25 T (ZFC state with ramp rate $\dot{B} = 0.03$ T/min), respectively. At $B = 0.1$ T, Fig. 4(a) shows that the $I$-$V$ characteristics overlap and exhibit identical features for the initial four runs (F1, R1, F2 & R2). However, in the F3 run, the driven vortices exhibit NDR transition, and the vortex matter enters the high $I_c$ state with $I_c^h = 46$ mA, followed by large fluctuations in voltage. The time series of these $V$ (or $u$) fluctuations around the jammed state near $I_c^h$ (data not being shown here) were similar to those reported earlier [26,27]. For all subsequent runs, we observe the vortex state depinning at $I_c^h$. Figure 4(b) shows that at 0.4 T, the vortex state depins from low $I_c$ state in F1and R1 runs and the high $I_c$ state is attained on F2 run. Figures 4(c), 4(d) show that the pristine vortex matter gets depinned at $I_c^h = 46.5$ mA, and we did not observe depinning from a low $I_c$ value at these relatively higher magnetic fields of 0.8 and 1.25 T.

Based on all $I$-$V$ data at various $B$, we count the number of $I$ cycles required to reach $I_c^h$ state via NDR-like transition as a function of $B$ (at 2.5 K). We denote these cycles as $N$ for each complete $I$-$V$, viz., $N = 1$ when the $I_c^h$ state is reached via the NDR transition in the first forward $I$-$V$ run, F1, or $N = 2$ when the $I_c^h$ state is reached via the quasi-NDR transition in the first reverse $I$-$V$ run, R1 and so on. The $N$ count is inversely proportional to the probability of observing the NDR (quasi-NDR) transition in the FF state. We treat NDR or quasi-NDR on same footing as they are associated with the same phenomenon. The main panel of Fig. 5 shows the behaviour of $N(B)$ for the ZFC state with $\dot{B} = 0.03$ T/min at 2.5 K. The ZFC state in Fig. 5 shows $N(B)$ decreases with increasing $B$ and $N(B) = 1$ for $B \geq 0.6$ T. This shows that in $2H$-NbS$_2$, the FF state at higher fields ($> 0.6$ T) is more susceptible to exhibit the NDR transition to produce the high $I_c$ state compared to that at lower $B$ for the same field sweep rate of 0.03 T/min. The inset panel of Fig. 5 shows that over the whole $B$-range (from 0.025 T to 1.25 T), $N(B)$ is almost uniform (at 4 cycles) for the FC vortex state, while the $N(B)$ decreases with $B$ (above 0.2 T) for the ZFC state. Thus, although the nature of $I$-$V$ curves is similar for the ZFC and FC states, the probability of observing the NDR transition is different for these two thermomagnetic history states. For our measurements, we also know that statistically, at any $B$, the probability of observing an NDR transition always has a non-zero finite value at any $\dot{B}$. A vortex state produced at a fixed $B$ which is reached with a high $\dot{B}$, the probability for observing the NDR transition in its $I$-$V$ is low (i.e., the $N(B)$ value is high), while the probability is high for a smaller $\dot{B}$. The finiteness of the probability for observing the NDR transition confirms that the high $I_c$ branch of the $I$-$V$ curve is a more stable one compared to the low $I_c$ branch. We would like to mention here that between samples taken from three different batches of $2H$-NbS$_2$ crystal, the difference in $I_c^l$ values between the different crystals is within about 5 mA [27]. In our paper by varying $B$, in the same crystal we are able to study the behaviour of $N(B)$ alongside wide variation in $I_c^l$, which also encompass different pinning regimes. In section 1 of Supplemental Material [36], we show the transformation into the high $I_c$ state from the low $I_c$ state by repeated $I$-$V$ cycling for a sample (A2) taken from a different batch. In section 5 of Supplemental Material [36], we show all $I$-$V$'s corresponding to the data shown in Fig. 5. In section 6 of Supplemental Material [36], we show that once the high $I_c$ state is reached after four $I$-$V$ cycles, this state is retained even with additional $I$-$V$ cyclings (five more $I$-$V$ cyclings). In section 7 of Supplemental Material [36], we show that over the entire $I$-$V$ measurement, the fluctuations in sample temperature are well below 5 mK.

## 4.  Discussion

Our Fig. 1(c) shows that most of the $B$ values, at which we perform our experiments, are well in the elastic vortex regime. As per the edge contamination model, for an elastic vortex state (ordered), a low $I_c$ state is reached if the disordered phase is injected into the sample from the edges at a rate (controlled by $\dot{B}$) slower than the annealing rate of the injected disorder in the elastic vortex state [34,35]. We have already shown in Fig. 3 that irrespective of the field sweeping rate, the driven vortex matter always transforms into a high $I_c$ state with repeated cyclings of current. From the edge contamination process [34,35], a fast field sweep rate triggers an annealing transformation of the injected disordered state into an ordered state in the elastic regime of the vortex state, viz., fast field sweep rates favour a high to low $I_c$ transformation [30,31]. However, here in $2H$-NbS$_2$, we see the opposite feature, viz., even with very fast field sweep rates there is a transformation from the low to high $I_c$ state. In fact, the high $I_c$ state is the more stable branch of the $I$-$V$ curve. All of these show that, although it is an ever-present process in all $I$-$V$ measurements, our results are not fully explained within the edge contamination process [34,35].

In the FF regime, the flux flow resistivity ($\rho_f$) is related to normal state resistivity ($\rho_n$) via, $\rho_f = \rho_n \frac{B}{B_{c2}}$ where $B_{c2}$ is the upper critical field [39]. Studies in high $T_c$ superconductors have shown that at low temperatures ($T \ll T_c$), using large pulsed currents, it is possible to attain flow resistivity, $\rho \sim 2\rho_f$ in FF regime. Such studies in HTSC and MoGe thin films [40,41,42,43] showed that at $\rho/\rho_f \sim 2$, an NDR transition occurred. These studies conclude that a reduction of the viscosity of the moving vortices leads to a rapid rise in vortex velocity. At large flow velocities, dynamic instabilities in driven vortex state lead to an NDR event [40,41,42,43,44,45,46,47]. However, it must be mentioned here that unlike the NDR transition in $2H$-NbS$_2$, no high $I_c^h$-like state was reported to have been reached via the NDR transition in HTSC. In $2H$-NbS$_2$, with $\rho_f = 60$ μΩ-cm, one finds that near the onset of NDR transition in the FF regime, $\rho/\rho_f \sim 7$ [see Fig. S1, section 8 of Supplemental Material [36]]. In $2H$-NbS$_2$, the high $\rho/\rho_f$ value is reached quite easily with comparatively much lower currents than in HTSCs [40-43]. We present here a possible scenario to explain the high $\rho/\rho_f$ values reached in $2H$-NbS$_2$. Local density of states studies using STM have shown localized bound states within the vortex core in $2H$-NbS$_2$ [9]. The absence of any underlying CDW in $2H$-NbS$_2$ (which modifies the spectrum of these quasi-particle bound states) perhaps favours more weakly localized quasi-particles within its vortex cores compared to other materials.

In the driven vortex state, Larkin-Ovchinnikov (LO) [48] proposed that the electric field acting across the vortex cores accelerates the quasi-particles in them. Once these bound quasi-particles acquire sufficient energy, they delocalize out of the vortex cores. In $2H$-NbS$_2$, due to the weakly bound nature of the quasi-particles in the vortex core, low drives may be sufficient to delocalize the quasi-particles. Due to this delocalization, LO [48] considered a non-thermal quasi-particle distribution around the vortex, which causes the vortex core to shrink, and hence a significant drop in the vortex viscosity. The conditions for attaining the non-thermal quasi-particle distribution is that the electron-electron scattering time, $\tau_{ee}$ is greater than the electron-phonon scattering time, $\tau_{ep}$. Using the expression for $\tau_{ee}$ and $\tau_{ep}$ in Ref. [42,49], we estimate for $2H$-NbS$_2$, $\tau_{ee}/\tau_{ep} = \frac{1}{r^2} \frac{E_F T}{k_B T_D^2} = \frac{7.73}{r^2} > 1$, where, the Fermi energy $E_F = 30$ eV [50], Debye temperature $T_D = 300$ K [51], $T = 2$ K and $r < 1$ is a dimensionless parameter related to phonon reflection coefficient at boundaries [42]. It is clear that for 2$H$-NbS$_2$, the value of $\tau_{ee} >> \tau_{ep}$ favours a shrink in vortex core which decreases the vortex viscosity, and hence significantly speeds up the vortices. The increased speed of the vortices results in enhanced flow resistivity. Recall from Fig. 1(b) in 2$H$-NbS$_2$, $u = 100$ cm/s near NDR, compared to $u \sim 1$ cm/s in 2$H$-NbSe$_2$ [52]. The rapid (fast $u$) injection of vortices from the irregular sample edges would trigger a rapid re-organization of vortices within the moving vortex state in the sample. The lowered viscosity results in unstable flow and the NDR transition [40-47] in 2$H$-NbS$_2$. The re-organization time-scale is of the order of, $\tau_{\text{re-org}} \sim \langle\delta\rangle / u$ , where $<\delta>$ is the mean spacing of vortices in the sample. As $\langle\delta\rangle \sim a_0 \propto 1/\sqrt{B}$ , hence $\tau_{\text{re-org}} \propto 1/\sqrt{B}$ . A decrease in $\tau_{\text{re-org}}$, at higher $B$ (for fixed $u$) or high $u$ (at fixed $B$), corresponds to a higher probability ($\propto 1/\tau_{\text{re-org}}$) for triggering rapid vortex re-organization in the flowing vortex state. This emerging dynamic instability triggers the NDR transition event (recall $N$ is inversely proportional to the probability of observing NDR).

For a static elastic vortex matter in a random pinning environment, the ZFC and FC vortex states have different spatial correlation volumes [22], which depends on the competition between the pinning strength and elasticity of the static vortex matter. If the average size of the correlation volume in the ZFC and FC vortex states are similar, then it is natural to expect that the $I_c$ and the $I$-$V$ curves for these states would be roughly similar. However, we find that although the $I$-$V$ curves are similar, the $N(B)$ curve is sensitive to differences in the ZFC and FC flowing states. This suggests that some subtle differences in the flowing vortex configuration exist deep in the FF regime. It is tempting to imagine an inhomogeneous FF state with channels of fast flowing uncorrelated vortices present alongside slowly drifting islands with relatively higher intervortex spatial correlations. A routine $I$-$V$ measurement is sensitive only to average $u$. However, inhomogeneity features affect $N(B)$ determined from $I$-$V$ cycling measurement. We argue that in the WP regime, the vortex configuration in the FF state created from the ZFC state is dominated by the slowly drifting islands. As this flow regime has significant regions with $\langle\delta\rangle \sim a_0$, the ZFC - $N(B)$ curve has a significant $B$ dependence (cf. Fig. 5). On the other hand, the FF regime of the FC state has more flow channels where $\langle\delta\rangle$ is not related to $a_0$, hence here the $N$ dependence on $B$ is weak. In fact, in the strong pinning (SP) regime, both ZFC and FC flow states are dominated by uncorrelated flow regions. Hence, in this regime as well, $N$ shows weak $B$ dependence (cf. Fig. 5). We believe that these differences in the inhomogeneity of the ZFC and FC flowing vortex states arise from subtle differences in the nature of the static vortex state prepared prior to depinning, which in turn is sensitive to the pinning regime (hence on $B$ and $T$ as well). We speculate some of these differences may arise from differences in distribution of topological defects (for example distribution of regions with five-seven fold co-ordinated vortices) while preparing the static vortex state. These differences may not leave a mark in the bulk $I_c$ values for ZFC and FC states, but they appear in the $B$ dependence of $N$. At high drives, the NDR (or quasi NDR) transition drives the traffic of flowing vortices jammed to a halt. This jammed state has high disorder with intervortex correlation lengths at least of the order of nearest neighbour distance. The $I_c$ of this state is $I_c^h$, which is much higher than that achievable by any conventional means. In samples with nominal pinning, due to the ever-present intervortex interactions the static vortex state prior to depinning doesn't usually achieve the maximally disordered vortex configuration (intervortex correlation is at least $\sim a_0$) when prepared through any conventional means like field cooling or accessing different pinning regime by varying $B$ and $T$. In a random pinning environment with interactions, one route to achieving a high entropy vortex configuration that is close to being maximally disordered, is via the NDR transition in the vortex flow state. In $2H$-NbS$_2$, depinning from $I_c^h$ is the limiting case of $I$-$V$ curves, as all curves with lower $I_c$ correspond to lower entropy vortex configurations. Considering these studies, it maybe also worthwhile recalling an earlier magnetization measurement in $2H$-NbSe$_2$ [31,53]. It was found that minor hysteresis loop measurements initiated from the FC vortex state in these crystals with moderate pinning resulted in newer critical current density states with magnetization values lying outside the envelop magnetization hysteresis loop. The envelope loop is typically rationalized by the critical state model [54] by assuming that the critical current is a single valued function of $B$. The multi-valuedness in field dependence of critical current exemplifies the higher critical current states in $2H$-NbSe$_2$, which were not accessible by usual routes of field cyclings starting from ZFC or FC modes (e.g., see Fig. 3 of Ref. [31]). Thus, the dynamic instability-driven NDR transition in the FF state of $2H$-NbS$_2$ is a route to achieve an unconventional high current density state. It is noteworthy that AC dynamics [31,55,56,57] produces different metastable pinned vortex configurations with different $I_c$'s. These AC dynamics experiments are performed with driving AC frequencies in the range of few 10's kHz and AC amplitudes of up to few 10's Oe. Compared to these AC frequencies the frequency of the repeated $I$-$V$ cyclings in this paper are orders of magnitude slower and the driving currents used are much higher. Here we would like to also emphasize that that the $I_c^h$ state reached isn't another metastable vortex pinned configuration with high $I_c$. The high $I_c^h$ state shows distinct features like distinct $I$-$V$ scaling properties around $I_c^h$, which are different from those around $I_c^l$ [25]. Also near $I_c^h$, we observe a critical slowing down of dynamics with features suggesting that the vortex state exhibits a non-equilibrium phase transition at $I_c^h$ [26]. It is interesting to note that AC dynamics experiments have also shown the presence of similar slowdown of dynamics suggestive of a non - equilibrium phase transition near the plastic depinning of vortex state in MoGe thin films [58]. It is an interesting prospect for future investigations to explore the comparison of the effects of AC dynamics on the non – equilibrium driven vortex states reached via transport measurements.

## 5. Conclusion

In conclusion, we have shown the protocol to produce a unique disordered high $I_c$ state from the conventional low $I_c$ state. This high $I_c$ state is not accessible by other means. We observe unusual features related to a very high velocity state in $2H$-NbS$_2$, which we believe is related to peculiar features related to the microscopic features of the vortex core in this system. Our results are interpreted in terms of delocalization of bound quasiparticles inside the vortex core in $2H$-NbS$_2$, which results in a drop of vortex viscosity at high drives. The consequent rapid increase in vortex velocity triggers reorganization in the flowing state leading to dynamical instability and the NDR transition. We hope that our experiments will motivate future explorations into these unusual non-equilibrium routes for reaching unusual high current density states.

**Acknowledgements**

S.S.B. acknowledges funding support from IITK (IN) and DST-AMT-TSDP (IN), DST-SERB Imprint II (IN), Government of India. AKS thanks DST for support under Year of Science Professorship. The authors would like to dedicate this work to Late Prof. A. K. Rastogi (retired from Jawaharlal Nehru University, New Delhi-110067, India). We also thank Late Prof. A. K. Rastogi and Prof. Asad Niazi (Jamia Millia Islamia (Central University), New Delhi - 110025, India) for the single crystals of $NbS_2$.

**References:-**

response for the jammed high Ic state, reached in R(B) in Fig. 2 (manuscript); (5) All I-V curves associated with Fig. 5 of the manuscript for ZFC and FC states; (6) Nine *I-V* cyclings. After 4 cycles (F1, R1, F2, R2), the high $I_c^h$ state is reached (in F3). After that further cycles F3, R3, F4, R4, F5 keeps the system in the high $I_c^h$ state; (7) Temperature very near the sample during I-V measurement; (8) Variation of dissipation with drive

# Supplemental Material:

Negative differential resistance state in the free - flux - flow regime of driven vortices in a single crystal of $2H$ -NbS$_2$

**Biplab Bag[1,+], Sourav M. Karan[1], Gorky Shaw[2], A. K. Sood[3], A. K. Grover[4] & S. S. Banerjee[1,*]**

[1] Department of Physics, Indian Institute of Technology, Kanpur, 208016, India
[2] National Physical Laboratory, Hampton Road, Teddington TW11 0LW, United Kingdom
[3] Department of Physics, Indian Institute of Science, Bengaluru, 560012, India
[4] Department of Applied Sciences, Punjab Engineering College, Chandigarh 160012, India
[+] Present Address: DCMPMS, Tata Institute of Fundamental Research, Mumbai, 400005, India

*email: satyajit@iitk.ac.in

1. **NDR seen in a sample from a different batch (Another Sample from different batch (A2 batch of Ref. 27, B. Bag, et al., Sci. Rep. 7, 5531 (2017)))**

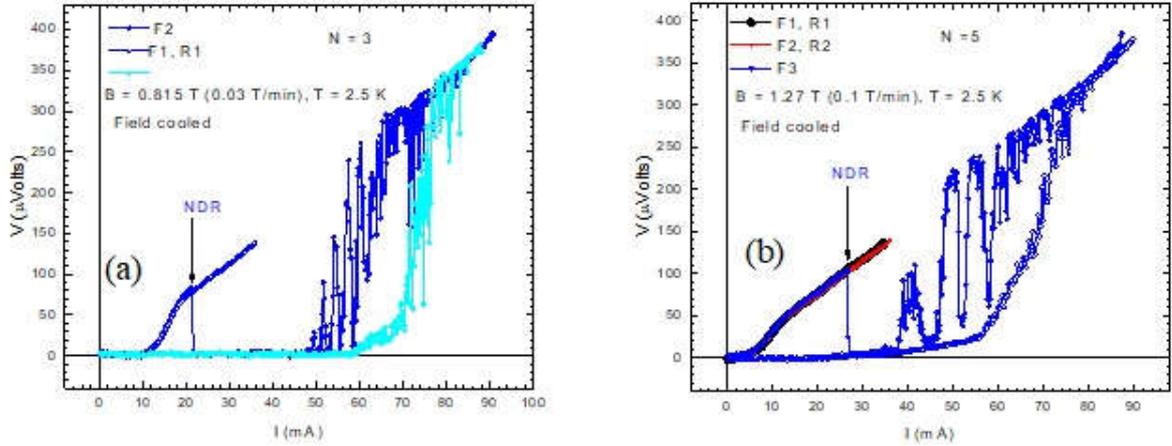

Figure S1: (a) shows onset of NDR after two cycling at 0.815 T (0.03 T/min) at 2.5 K, FC state. (b) shows onset of NDR after 4 cycling at 1.27 T (0.1 T/min) at 2.5 K, FC state.

From Fig. S1, we see that in this sample (say A2), at 0.815 T and 1.27 T, NDR transition occurs for $N = 3$ and $N = 5$ respectively. Whereas for the sample (say A1) in the present manuscript, in this field regime, $N = 4$. Hence, A2 sample takes one more/less cycle than that in A1. We would like to state that while there maybe slight differences in the absolute value of $N$ from sample to sample, the overall nature of the phenomenon remains unchanged.

## 2. Structure and composition information

Powder XRD for samples from three different crystal batches, with their indexed peaks and Lattice parameters

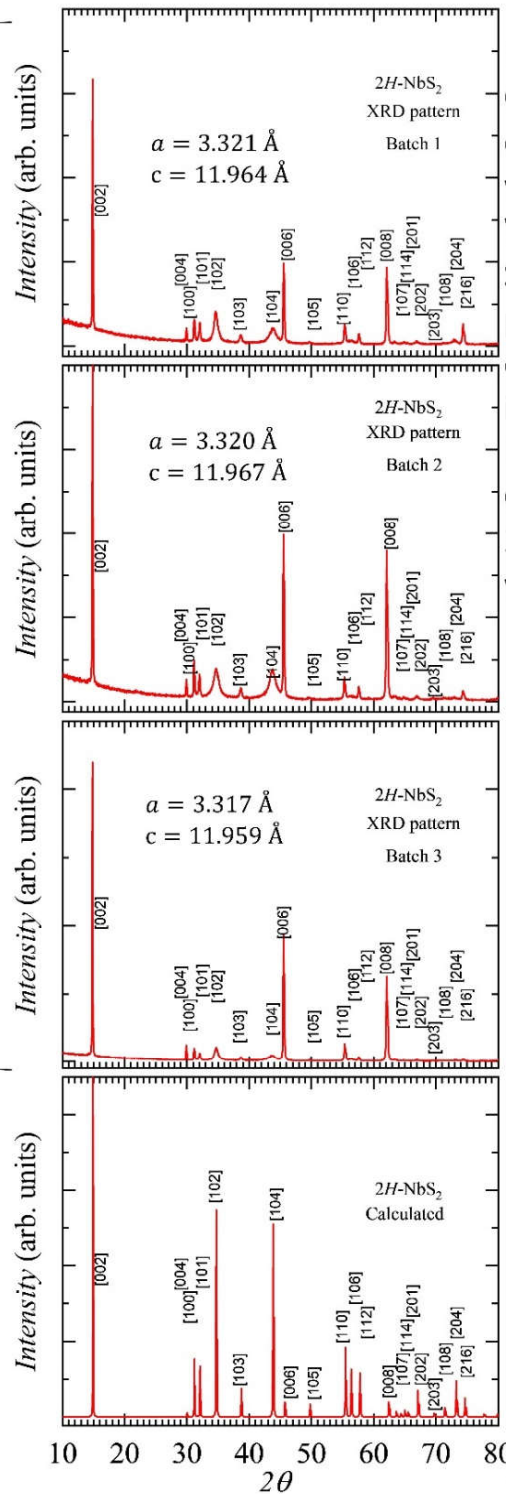

Powder XRD Analysis of Crystals taken from different batches along with their Indexing. We note that the crystalline Structure / lattice parameters, across the different crystal batches is almost uniform. The Lattice parameters Correspond with the 2H phase of NbS2 and we find no 3R phase.

Expected (calculated) peak locations for 2H-NbS$_2$. The actual Peaks closely match with the expected peak Locations.

**EDX: Atomic percentage in 2*H*-NbS$_2$ samples**

Atomic percentage in 2*H*-NbS$_2$ samples

| Batch | Nb (%) | S (%) |
|:-----:|:------:|:-----:|
| 1 | 34.18 | 65.82 |
| 2 | 34.78 | 65.22 |
| 3 | 34.63 | 65.37 |

EDX analysis was done at different locations on samples from three different batches. The average values are shown in above table. From the analysis, we see that on the average, the atomic ratio, S:Nb > 1.85. This is expected for the 2*H*-NbS$_2$ compound. The presence of a 3R phase would lead to a much lower S concentration (S:Nb < 1.8). The major components detected by EDX were Nb and S. Other than Nb and S the EBX software showed few other trace elements. The accuracy of the percentages of the minor quantity of other trace elements suggested by the EDX software, isn't certain. As their contributions was less than 0.05% hence we include them within the error bars of 0.05% the above values.

A common issue associated with 2*H*-NbS$_2$ is that if the growth conditions aren't carefully controlled, then it leads to the formation of the metal rich 3R phase, which is non-superconducting. The presence of a 3R phase in 2*H*-NbS$_2$ would act as a strong pinning center in the crystal. We show via powder XRD analysis that the lattice parameters as well as the location of the XRD peaks suggest the presence of only 2*H* phase and there is no evidence of the 3R phase in our crystals. The energy dispersive X-ray analysis (EDX) analysis also shows the absence of any 3R phase in the crystals.

### 3. *T*-dependence of $I_c^h$ & $I_c^l$; and demonstration of recovering back low $I_c^l$ state via field cooling the sample

For comparison, in Figs. S2(a),(b), we present the *T*-dependence of critical currents for the sample on which data is presented in the main manuscript. We find that $I_c^h$ is *T*-independent while in the same *T* range $I_c^l$ exhibits conventional decrease with increasing *T*. This shows the behaviour of $I_c^h$ and $I_c^l$ is quite distinct. Once one enters the $I_c^h$ state, we find that we can recover back the $I_c^l$ state only via going above $T_c$ and then field cooling down and measuring the *I-V* at the desired *B*, as shown in the next figure S3.

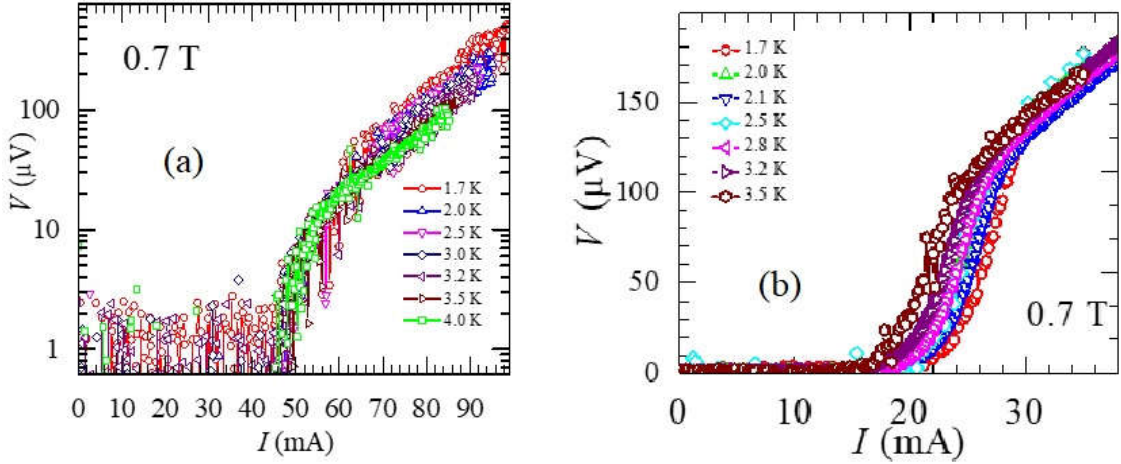

Figure S2: (a) *I-V* data at various *T*'s showing depinning curves across $I_c^h$. (b) *I-V* data at various *T*'s showing depinning curves across $I_c^l$.

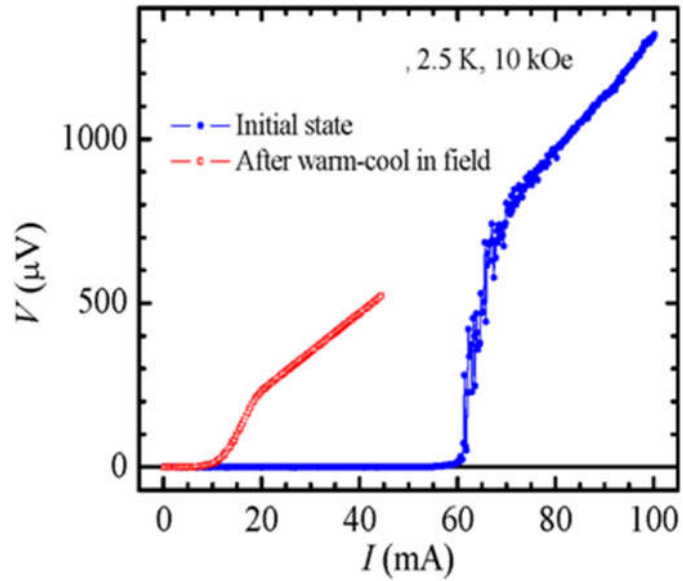

Figure S3: *I-V* response at 2.5 K and 10 kOe of a 2*H*-NbS$_2$ sample of different batch. The initial state belongs to a zero field cooled vortex state and the after warm-cool state in field corresponds to a field cooled (FC) state

The *I-V* data shown in Fig S3 is measured in the crystal from as different batch of 2$H$-NbS$_2$ sample. The curve marked as "Initial state", shows the *I-V* corresponding to the high $I_c^h \sim 60$ mA state attained by *I-V* cycling. Next, without changing the applied magnetic field, we switch off the applied current and warm the sample to 20 K which is well above its $T_c$ (5.8 K), and then cool down to 2.5 K in the applied field of 10 kOe (field cooling). In this field cooled (FC) state, as we measure the *I-V* of the sample, we observe that for this FC state, the driven vortex matter depins from the low $I_c^l$ state. Had regions of structural and/or chemical inhomogeneous regions which act as strong pinning centers which could be responsible for the high $I_c$ state, then upon field cooling we should have obtained the high $I_c$ state rather than the low $I_c$ state, as vortices should have been preferentially pinned on these strong pinning centers.

### 4. *I-V* response for the jammed high $I_c$ state, reached in $R(B)$ in Fig. 2 (manuscript)

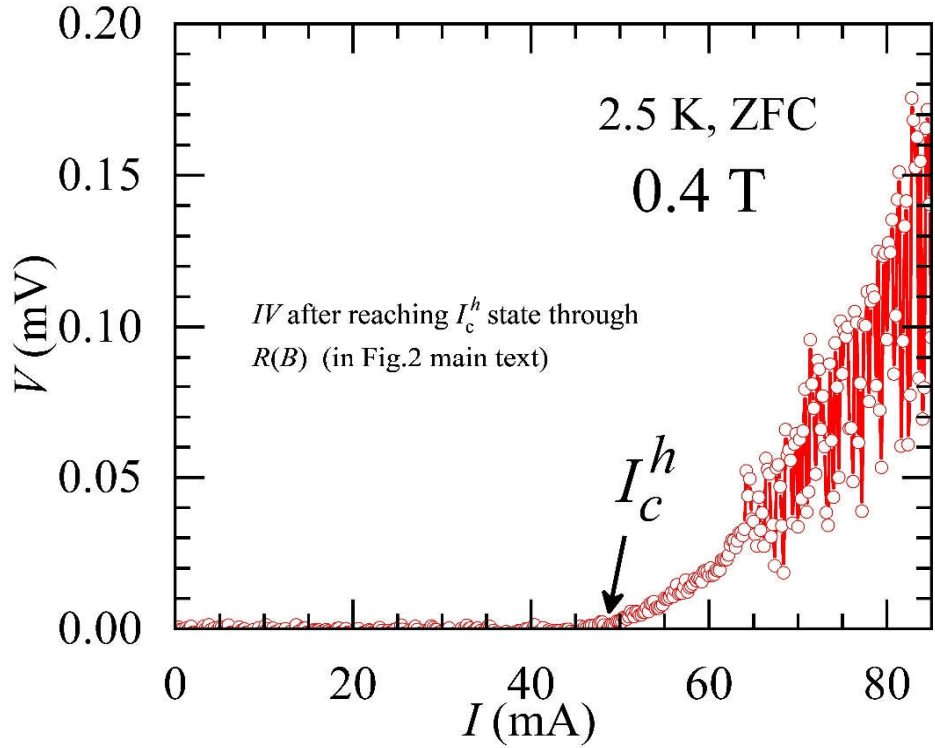

Figure S4: *I-V* response at 2.5 K and 0.4 T (ZFC) for the high $I_c$ state reached via $R(B)$ in Figure 2 (manuscript)

After transformation from the low to high $I_c$ state, reached in $R(B)$ measurements (Fig. 2 of manuscript), we measure the *I-V* in that high $I_c$ vortex state (Fig. S4), which exhibits depinning at $I_c^h \sim 45$ mA.

**5. All *I-V* curves associated with Fig. 5 of the manuscript for ZFC and FC states**

ZFC data

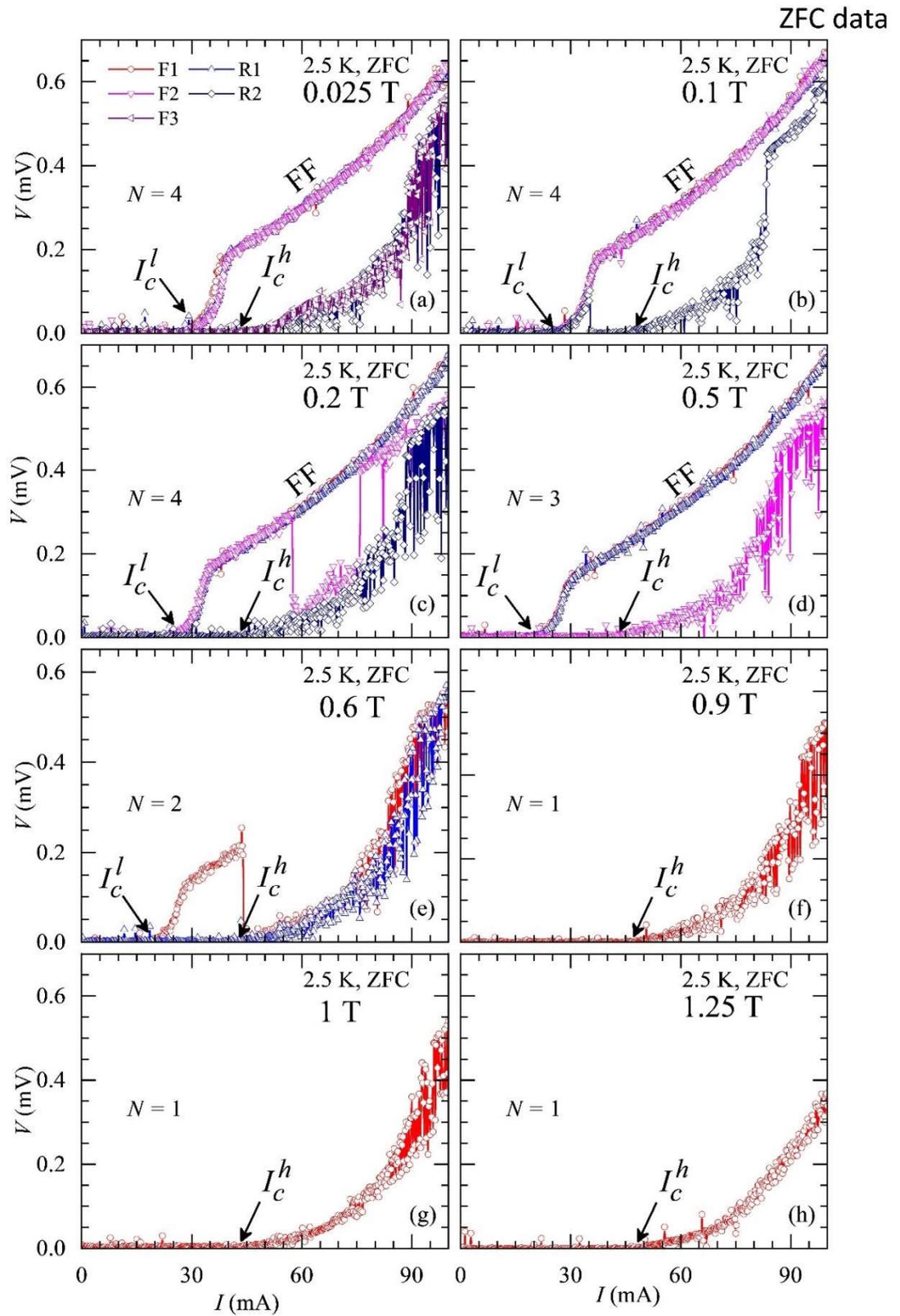

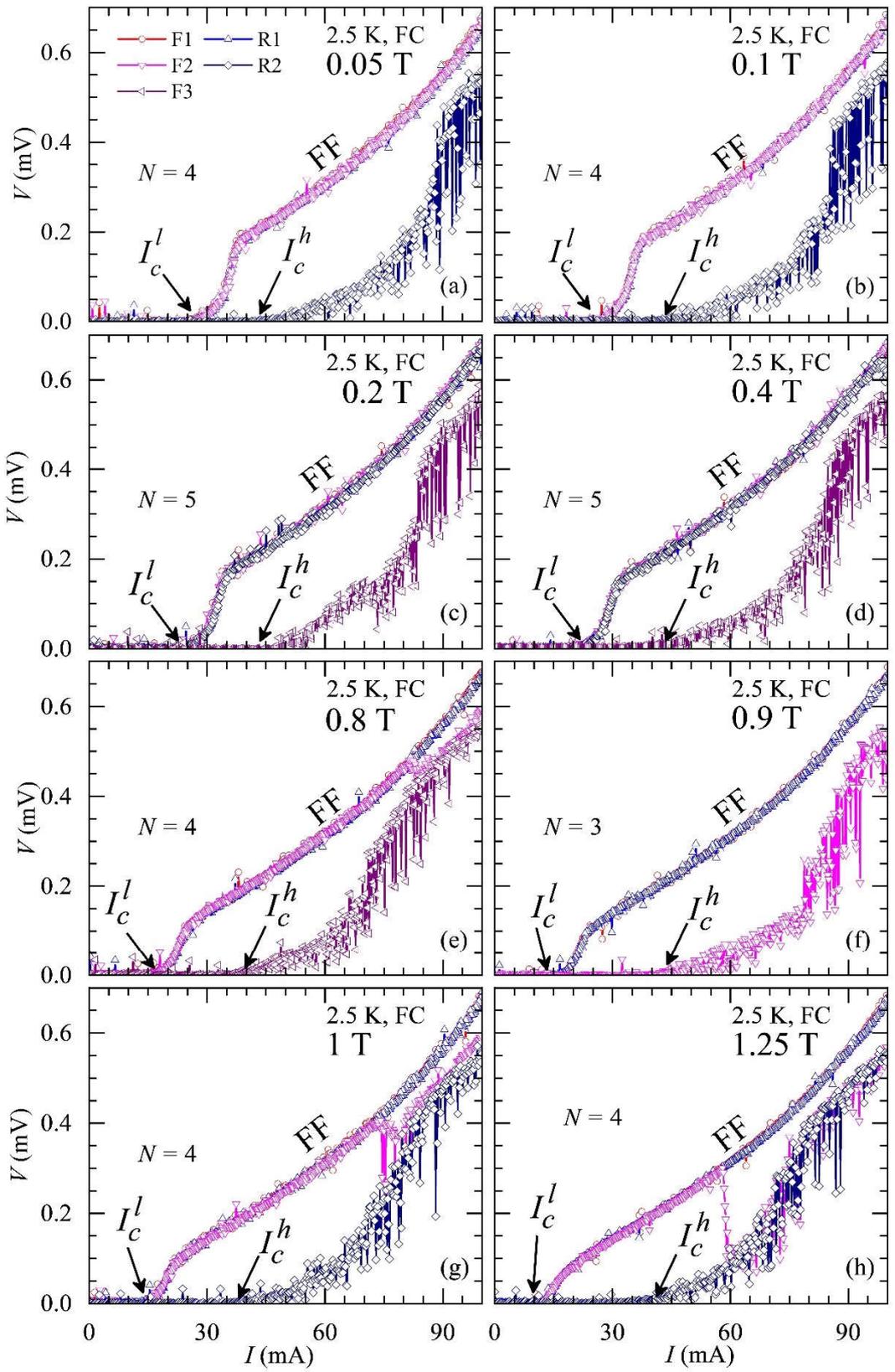

6. **Nine *I-V* cyclings. After 4 cycles (F1, R1, F2, R2), the high $I_c^h$ state is reached (in F3). After that further cycles F3, R3, F4, R4, F5 keeps the system in the high $I_c^h$ state.**

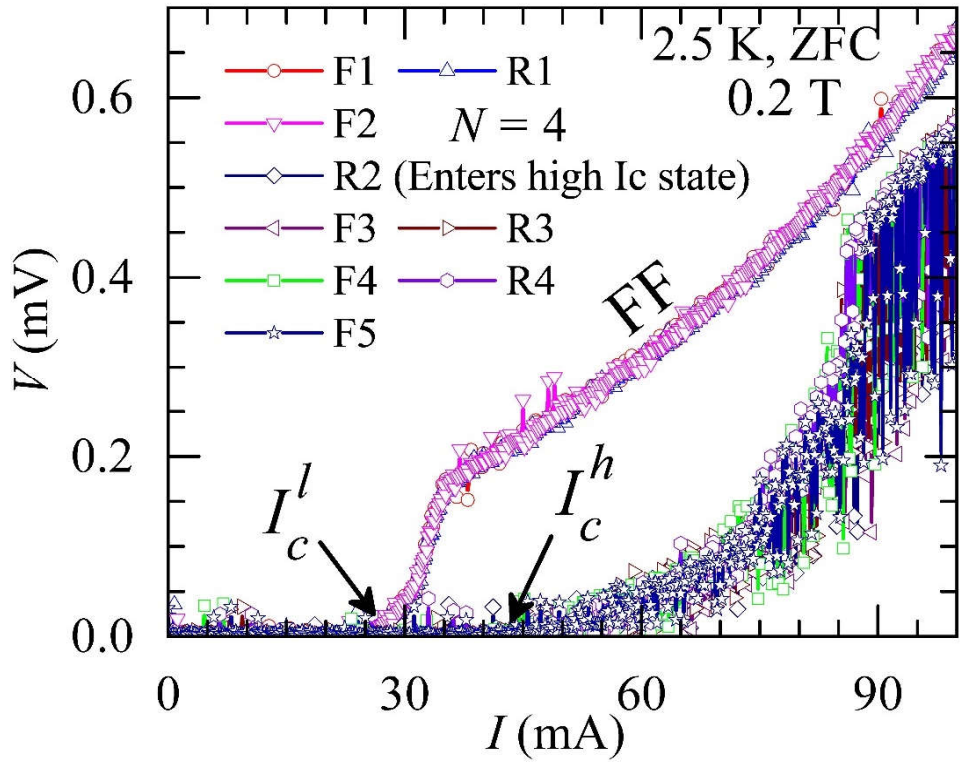

## 7.  Temperature very near the sample during *I-V* measurement.

Temperatures as measured by the temperature sensorAt 2.5 K, when the current(I) is ramped up to 100 mA.

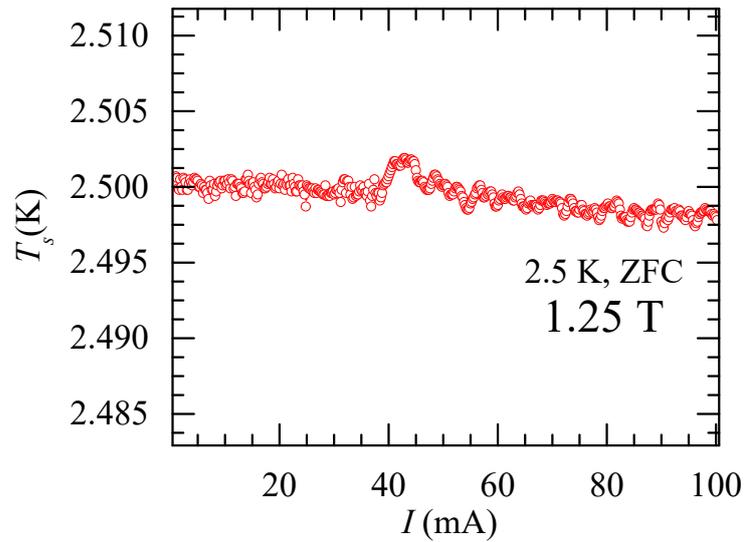

Figure S5: Variation of sample temperature while increasing *I* during *I-V* measurement at 2.5 K and 1.25 T

We show a measurement of sample temperature ($T_s$) as recorded by the temperature sensor located ~ 1.0 mm away from the sample, at 2.5 K. This $T_s(I)$ measurement is performed, during the *I-V* measurement as current is increased upto 100 mA. Note that that most of the fluctuations in $T_s$ are well within 5 mK during the *I-V* measurements.

## 8. Variation of dissipation with drive

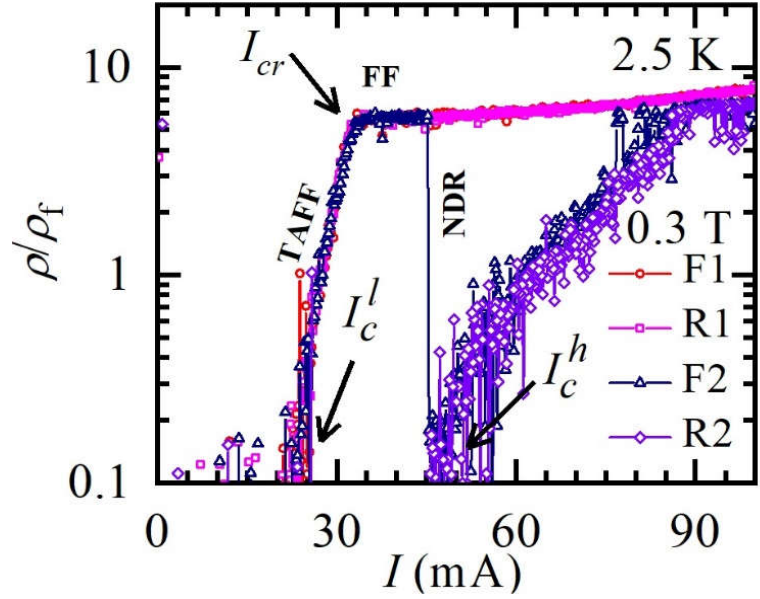

Figure S6: $\rho/\rho_f$ vs $I$ (in log linear scale) at 2.5 K and 0.3 T for F1, R1, F2 and R2 runs.

In figure S6, we have replotted the *I-V* data in Figs. 1(a),(b) in terms of normalized resistivity $\left(\rho/\rho_f\right.$, where $\rho$:resistivity and $\rho_f$: Bardeen-Stephen flux flow resistivity$\left.\right)$ as a function of $I$ where $\rho_f = \rho_n \frac{B}{B_{c2}}$, $\rho_n = 60$ μΩ-cm (normal state resistivity) and higher critical magnetic field, $B_{c2} = 2.5$ T (at 2.5 K). It shows that for the F1, R1 and F2 (for details see the manuscript) after depinning from the pristine vortex state at $I_c^l \sim 23$ mA, we observe sharp enhancement in dissipation ($\rho$) with drive ($I$) in the thermally activated flux flow regime (TAFF) and the driven vortex state transforms into the flux flow (FF) state beyond $I_{cr}$. The most noteworthy feature in Fig. S6 is that the dissipation is almost constant ($\rho/\rho_f \sim 7$) in the FF regime for F1, R1 and F2 runs. Additionally, in the FF regime the driven vortex matter for the F2 run exhibits NDR transition (at $I \sim 45$ mA, for details see the main text) and falls to a drive-induced immobile state (high $I_c$ state) which depins at $I_c^h$, following a noisy response in *IV* or $\rho/\rho_f(I)$. However, the dissipation levels in this fluctuating levels are considerably small compared to the FF state. This fluctuating dissipation response turns uniform and merges with the FF state beyond $I \sim 80$ mA.